# The Inward Dispersion of the Neutron Scattering Experiments in HTSC Cuprates.

Moshe Dayan

Department of Physics, Ben-Gurion University,
Beer-Sheva 84105, Israel.

E-mail: mdayan@bgu.ac.il

Key Words: High Temperature Superconductivity.

# The Inward Dispersion of the Neutron Scattering Experiments in HTSC Cuprates.

Moshe Dayan

Department of Physics, Ben-Gurion University,
Beer-Sheva 84105, Israel.

## Abstract

The theory of the high temperature superconducting cuprates, which is based on the condensation of holes into strings in checker-board geometry, was successful to explain the elastically scattered Neutrons by spin waves. Here it is extended to analyze the inward dispersion curve of its inelastic counterpart, up to the resonance energy- $E_{res}$. This extension is done by applying the perturbation theory of the linear response to the condensed strings. The approximated susceptibility is derived by means of the ring diagram. The dispersion relation is obtained from the dispersion of the poles of the susceptibility integral. It is found that the particle anti-particle pair that yields the susceptibility is the time reversal pair where the particle momentum is $k$ in phase A, and the anti-particle momentum is $-k$ in phase B. The dispersion is found to be in agreement with experiment, subject to some suggested corrections. The weak intensity by the resonance energy, as well as the dispersion, is speculated to be modified due to interference with spin waves that are caused by direct spin flip, as in the mother undoped materials.

## 1. Introduction

Neutron scattering experiments (NSE) are undoubtedly one of the most important tools for investigating the magnetic ordering of high temperature Copper-Oxide Superconductors, and therefore for revealing the origin of its pseudo-gap state. The undoped "mother-materials", in each Cuprate family, exhibit Anti-Ferromagnetism (AF), which is manifested by scattered peaks at the reciprocal lattice vector- $Q_{AF} = \frac{2\pi}{a}(0.5, 0.5)$. Here the lattice parameter- a is the two dimensional lattice parameter of the $CuO_2$ planes, which is approximated by its tetragonal value (even when the structure is slightly orthorhombic). The scattered peak indicates the doubling of the area of the lattice cell, due to the AF order. When the cuprates are doped in the underdoped regime, namely enough to be electrical conductors, but still below their maximum superconductivity, the AF peak at $Q_{AF}$ disappears and is replaced by four incommensurate peaks. At low energies, these incommensurate peaks appear at the reciprocal lattice vectors: $\frac{2\pi}{a}(0.5 \pm \delta, 0.5)$, and $\frac{2\pi}{a}(0.5, 0.5 \pm \delta)$, where $\delta$ is roughly equal to the doping fraction x [1, 2]. These peaks were interpreted, in my former paper [2], to be originated from spin density waves (SDW), with amplitudes that are proportional to the order parameter of the pseudo-gap state. The doped holes, that produce these SDW, are assumed to aggregate into linear strings in the x and y directions. The latter are located between two different AF phases (phase A and phase B), so that their movements change the borders between these phases, while keeping their order intact. The ground state is a superposition of all such states, in which the strings occupy combinations of positions. The ground state is built up to include a SDW component that makes up the internal field which guarantees the ordered pseudogap state.

The former conventional perception was that the incommensurate peaks are observable only in the normal state. In the superconductive state they appear only at energies larger than some spin-gap [1, 3-5]. More recently, this perception has been doubted and even contradicted [27]. Anyway, when energy is increased above the superconductive gap, the Neutron scattered beams exhibit inward dispersion with respect to reciprocal lattice momentum [1, 3-6], until the four peaks merge into a single one at $Q_{AF}$. The energy at which this occurs is usually referred to as the "resonance energy"- $E_{res}$. Above the resonance energy the dispersion is outward, and the peaks appear at $\frac{2\pi}{a}(0.5 \pm \alpha(\omega), 0.5 \pm \alpha(\omega))$, where $\alpha(\omega)$ is an increasing function of $\omega$, according to Hayden et al. [7], or at circular positions around $Q_{AF}$, according to Stock et al. [8].

Tranquada included various NSE data from different investigations, on different materials, into a single figure for comparison [1]. This figure is reproduced and shown here in Fig. 1. Notice that the energy scale is normalized by the super-exchange magnetic parameter $J$. The overall Hour-glass shape of the data is conspicuous. This includes the low energy inward dispersion which converges to $E_{res}$, and the high energy outward dispersion.

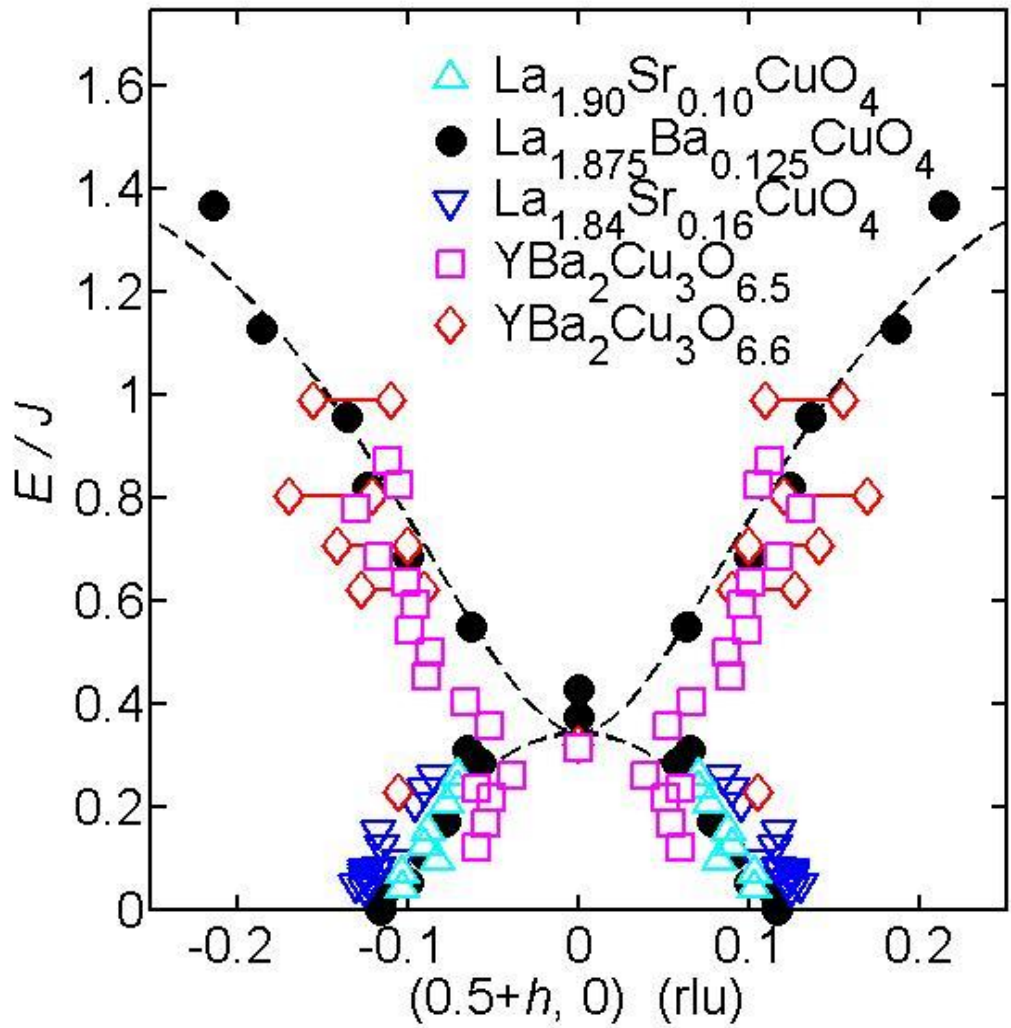

**Figure 1**. The dispersion of the incommensurate spin waves from various experiments on LSCO, LBCO, and YBCO samples, on a mutual scale . The energy scale is normalized by the super-exchange magnetic parameter $J$. Reproduced from Tranquada [1].

The outline of the present paper is the following. First I briefly present the theoretical method, which is based on the theory of linear response of the system to the magnetic perturbation of the neutron beam. This naturally introduces the approximated spin susceptibility of the strings of holes. Since this approximation is based on the so-called "ring diagram" of a particle-antiparticle pair, we reach the question of the momentum and energy relationships between the particle and the antiparticle. In the present paper, a pedagogical approach is preferred. Since it is shown in [2] that the pair $(k, \bar{k})$ is responsible for the elastic peaks, it is reasonable to assume such pairs and check whether they produce the experimental dispersion. The zero dispersion that is associated with the former definition of $\bar{k}(k)$ [2], suggests the revised definition of the forthcoming Eq. (18a,b). This revised definition results in an inward dispersion, but it is shown, that it is much smaller than the experimental one. Finally, we arrive to the proposal that the pair $(k, -k)$ is the one that produces the experimental dispersion. The comparison with experiment is done in the last section, where also the reasons for the differences with the present theory are qualitatively discussed.

## 2. The low energy inward dispersion.

In [2], I analyzed the expectation value of the number operator of the column of holes - $C_j^\dagger C_j$, at the ground state, and found its intensity and its momentum dependence. Instead of generalizing that method to fit the present analysis, I would rather use the more systematic formalism, which is briefly introduced below. This is the formalism of linear response to the magnetic field of the Neutrons in NSE, as derived from the perturbation field theory [9-11]. The cross section for scattering a neutron beam of initial energy and momentum - $E_i, Q_i$, and final energy and momentum - $E_f, Q_f$, respectively, into a differential angular element $d\Omega$, and a differential energy $dE_f$, is

$$\frac{d^2\sigma}{d\Omega dE_f} = \frac{Q_f}{Q_i} [0.5 \gamma r_0 g F(\mathrm{q})]^2 S(\mathrm{q}, \omega).$$  (1a)

Here $\gamma$, g, and $r_0$ are physical constants representing the gyroscopic ratio, the Lande factor, and the classical radius of the electron, respectively. The function $F(\mathbf{q})$ is the magnetic form factor of a single scatterer [11]. All these later quantities could be lumped into one single parameter $C'$, to yield

$$\frac{d^2\sigma}{d\Omega dE_f} = C' \frac{Q_f}{Q_i} S(\mathbf{q}, \omega). \qquad (1b)$$

The structure factor is actually a tensor- $S^{\alpha\beta}(q, \omega)$, but here it is summed according to the following selection rule to result in the dynamical structure factor

$$S(\mathbf{q}, \omega) = \sum_{\alpha, \beta} (\delta_{\alpha, \beta} - q_\alpha q_\beta / \mathbf{q}^2) S^{\alpha\beta}(\mathbf{q}, \omega). \qquad (2)$$

Eq.(2) shows that only the spin components that are perpendicular to the scattering vector $q$ contribute. For our purposes, where we actually treat semi-one dimensional problem of columns of holes, $q$ is in the spin-wave direction and we simply use the scalar version of the structure factor. Now, at finite temperatures, according to the fluctuation dissipation theorem, the magnetic scattering factor is related to the general susceptibility $\chi(\mathbf{q}, \omega)$ by [1,11]

$$S(\mathbf{q}, \omega) = \frac{\hbar}{\pi} \operatorname{Im} \chi(\mathbf{q}, \omega) [1 - \exp(-\hbar\omega / kT)]^{-1}. \qquad (3)$$

Therefore, Eq. (1b) becomes,

$$\frac{d^2\sigma}{d\Omega dE_f} = C \frac{Q_f}{Q_i} \operatorname{Im} \chi(\mathbf{q}, \omega) [1 - \exp(-\hbar\omega / kT)]^{-1}, \qquad (4)$$

where $C$ is a constant. In the following we use units for which $\hbar = 1$.

The susceptibility $\chi(q, \omega)$ expresses the linear response of the magnetic system of the cuprates to the magnetic perturbation of the neutrons. In general, the simplest Feynman diagram for such a response depicts a bare particle-antiparticle excitation, which hereafter is referred to as a "ring diagram" [9, 10]. It is the simplest form of all irreducible diagrams. The ring diagram is a (fixed $q$ and $\omega$) integral over all the excitations from below the Fermi level to above it. The static part of the NSE, for non-superconducting samples, suggests that the above excitations have no energy gap. This, in turn, suggests the use of the gapless field $\psi_{k-}(t)$ of Eq. (35a) of [2], and its related excitation energy $E_{k-}$

$$E_{k-} = \sqrt{\varepsilon_k^2 + \Lambda_k^2} - |\Lambda_k| = E_k - |\Lambda_k|. \tag{5}$$

The related propagator of this field is

$$G_0(k, \omega) = \frac{1}{2} \left[ \frac{I + (w_k^2 - v_k^2)\tau_3 - 2w_k v_k \tau_1}{\omega - E_{k-} + i\delta\omega} + \frac{I - (w_k^2 - v_k^2)\tau_3 - 2w_k v_k \tau_1}{\omega + E_{k-} + i\delta\omega} \right]. \tag{6}$$

Eq.(6) is equal to Eq.(37) of [2], where all of its variables are defined. The first term represents particle-like excitations, and the second one represents antiparticle-like excitations.

To differentiate between the accurate susceptibility $\chi(q, \omega)$, and the susceptibility obtained from the ring diagram, we denote the latter one by $\Pi(q, \omega)$. This approximated susceptibility may also be expanded (by Dyson's equation, for the Random Phase Approximation [9, 10]) to define

$$\tilde{\Pi}(q, \omega) = \Pi(q, \omega)[1 - V(q, \omega)\tilde{\Pi}(q, \omega)], \tag{7a}$$

which yields for $\tilde{\Pi}(q, \omega)$ the solution,

$$\tilde{\Pi}(q, \omega) = \Pi(q, \omega)[1 + \Pi(q, \omega)V(q, \omega)]^{-1} , \tag{7b}$$

In Eqs.(7a,b), $V(q,\omega)$ is some interaction that links between indefinitely repeated ring diagrams. It should be remarked, though, that this interaction cannot be the magnetic interaction of Eq.(5) of [12], since the later interaction is limited within the same magnetic phase (either A or B), whereas $\Pi(q,\omega)$ involves propagators from two different magnetic phases. Also, assuming $V(q,\omega)$ to be the magnetic interaction between the spin system and the neutrons, $\tilde{\Pi}(q,\omega)$ violates the condition that the response is linear. This proves that our approximation of replacing $\chi(q,\omega)$ by $\Pi(q,\omega)$ is a reasonable one, subjected to the following remark. Here we use the propagator of Eq. (6) to calculate $\Pi(q,\omega)$, which implies that $\Pi(q,\omega)$ is the result of perfect strings of holes. It was proved that the actual situation is different, and imperfect strings are essential [12]. This limitation is qualitatively discussed in the next section.

The susceptibility $\Pi(q,\omega)$ is given by

$$\Pi(q,\omega) = \frac{i}{(2\pi)^3} \int d\nu\, d^2k\, G_0(k,\nu) G_0(k+q,\nu+\omega).$$ 
(8)

The product of the propagators in the integrand includes four terms. Two of these terms have two poles on the same side of the energy axis, and therefore vanish upon the energy integration. The rest two are

$$\Pi(q,\omega) = \frac{i}{(2\pi)^3} \int d\nu\, d^2k \{ \frac{\tilde{G}_p(k)\tilde{G}_{ap}(k+q)}{(\nu - E_{k-} + i\delta\nu)[\nu + \omega + E_{(k+q)-} + i\delta(\nu+\omega)]}$$

$$+ \frac{\tilde{G}_{ap}(k)\tilde{G}_p(k+q)}{(\nu + E_{k-} + i\delta\nu)[\nu + \omega - E_{(k+q)-} + i\delta(\nu+\omega)]} \}.$$
(9)

In Eq. (9)

$$\tilde{G}_p(k) = \frac{1}{2}[I + (w_k^2 - v_k^2)\tau_3 - 2w_k v_k \tau_1]. \tag{10a}$$

$$\tilde{G}_{ap}(k) = \frac{1}{2}[I - (w_k^2 - v_k^2)\tau_3 - 2w_k v_k \tau_1]. \tag{10b}$$

Both of the remaining terms in Eq. (9) have poles on both sides of the energy integration axis, and are therefore finite. The energy integration yields

$$\Pi(q,\omega) = \frac{1}{(2\pi)^2} \int d^2k \{ \frac{\tilde{G}_p(k)\tilde{G}_{ap}(k+q)}{\omega + E_{k-} + E_{(k+q)-} - i\delta} - \frac{\tilde{G}_{ap}(k)\tilde{G}_p(k+q)}{\omega - E_{k-} - E_{(k+q)-} + i\delta} \}. \tag{11}$$

Since $E_{k-} + E_{(k+q)-} \geq 0$, only the second term has poles for non-negative $\omega$, which correspond to processes in which the system extracts energy from the incoming neutron beam. Lets denote this term by $\Pi_+(q,\omega)$

$$\Pi_+(q,\omega) = \frac{1}{(2\pi)^2} \int d^2k \frac{\tilde{G}_{ap}(k)\tilde{G}_p(k+q)}{E_{k-} + E_{(k+q)-} - \omega - i\delta}. \tag{12}$$

Notice that the susceptibility $\Pi_+(q,\omega)$ should be causal. This implies the following relations with respect to the time ordered polarization $\Pi(q,\omega)$: $\mathrm{Re}\,\Pi_+(q,\omega) = \mathrm{Re}\,\Pi(q,\omega)$; $\mathrm{Im}\,\Pi_+(q,\omega) = \mathrm{Im}\,\Pi(q,\omega)\mathrm{sign}(\omega)$, which is fulfilled by our choice of $\Pi_+(q,\omega)$. We are interested meanwhile only in $k+q$ in close vicinity to $\bar{k}$. To be more specific, we are interested in $k+q = \bar{k} + \delta q$, with $|\delta q| << k_F$. Although the choice of $(k, \bar{k})$ as the "(string) particle anti-particle" pair will soon be found as non-dispersive, it is pedagogically worthwhile to start with it for the following reasons: 1) It seems as a natural starting point to continue our former treatment of the elastic limit [2], and possibly use it to approximate low energy inelastic processes. 2) It is a convenient preface for proposing a different definition of $\bar{k}(k)$, and checking its consequences, which is done later. 3) Some features of the resultant susceptibility are relevant also for the final choice of the "particle anti-particle" pair. Now, since for small $\varepsilon_k, \varepsilon_k \approx Ja^2(k^2 - k_F^2) \approx -\varepsilon_{\bar{k}}$, we have

$E_{(\bar{k}+\delta q)-} = E_{(k+\delta q)-} \cong E_{k-} + \delta q \dfrac{\partial E_{k-}}{\partial k}$. Thus, **for small** $|\varepsilon_k|$ **we approximate**

$$E_{k-} + E_{(\bar{k}+\delta q)-} \cong (\varepsilon_k^2 + 2\varepsilon_k Ja^2 k_F^2 \frac{\delta q}{k_F}) / |\Lambda_k|. \tag{13}$$

After performing the integration along the string direction, and changing the integration variable to $\int d\varepsilon_k$, we get

$$\Pi_+(q,\omega) = \frac{|\Lambda_k|}{4\pi\sqrt{J}} \int\limits_{-Ja^2 k_F^2 + \delta}^{3Ja^2 k_F^2} \frac{d\varepsilon}{\sqrt{Ja^2 k_F^2 + \varepsilon}} \frac{\tilde{G}_{ap}(k)\tilde{G}_p(\bar{k}+\delta q)}{\varepsilon^2 + 2\varepsilon a^2 k_F^2 J(\delta q / k_F) - \omega|\Lambda_k| - i\delta}. \tag{14}$$

The poles are $\varepsilon_\pm = -Ja^2 k_F^2 \dfrac{\delta q}{k_F} \pm [\sqrt{(Ja^2 k_F^2)^2 (\frac{\delta q}{k_F})^2 + \omega|\Lambda_k|} + i\delta]$, but only $\varepsilon_-$ is

relevant for $k$ of anti-particle. Integration yields the approximate result

$$\Pi_+(q,\omega) = i\frac{\Lambda_-}{4Jak_F} \frac{\tilde{G}_{ap}(\varepsilon_-)\tilde{G}_p(\varepsilon_+)}{\sqrt{(Ja^2 k_F^2)^2 (\frac{\delta q}{k_F})^2 + \omega\Lambda_-}}. \tag{15}$$

In Eq.(15), $\Lambda_-$ is the absolute value of the pseudogap at $k = k(\varepsilon_-)$. The matrix $\tilde{G}_{ap}(\varepsilon_-)\tilde{G}_p(\varepsilon_+)$ has diagonal and non-diagonal components. Their interpretation is discussed later. Meanwhile, we assume that the relevant component of $\tilde{G}_{ap}(\varepsilon_-)\tilde{G}_p(\varepsilon_+)$ is the $I$ component, which equals ½. For this component we have

$$\text{Im}\,\Pi_+(q,\omega) = \frac{\Lambda_-}{8Jak_F} \frac{1}{\sqrt{(Ja^2 k_F^2)^2 (\frac{\delta q}{k_F})^2 + \omega\Lambda_-}}. \tag{16a}$$

For fixed $\omega$, the susceptibility has its maximum when $\delta q = 0$, and

$$\text{Im}\,\Pi_+(q,\omega) = \frac{1}{8Jak_F}\sqrt{\frac{\Lambda_-}{\omega}} \;. \qquad (16b)$$

The divergence of the susceptibility at zero loss-energy, as shown by Eq. (16b), is not physical since there probably is a finite lifetime associated with the strings of holes, and also with the excitations, in the pseudogap state. Although this has not been studies so far, it probably should remove the said divergence. In the following I demonstrate this by assuming a simple model, which should not be taken too seriously. I represent the finite life times by a constant rate $\Gamma$, by simply making the transformation $\omega \to \omega + i\Gamma$ in Eq. (16b). Let us also define $\Omega = \sqrt{\omega^2 + \Gamma^2}$. This yields

$$\text{Im}\,\Pi_+(q,\omega+i\Gamma) = \frac{1}{8Jak_F}\sqrt{\frac{\Lambda_-(\Omega+\omega)}{2\Omega^2}} \;. \qquad (16c)$$

Now, at $\omega = 0$, $\Omega = \Gamma$, and $\text{Im}\,\Pi_+(q,\omega+i\Gamma) = \frac{1}{8Jak_F}\sqrt{\frac{\Lambda_-}{2\Gamma}}$, which is finite. Another way to avoid the static limit diversion is given by introducing a small low energy limit for the treatment of the present paper. This is done in the forthcoming Eq.(26). However, the fundamental way to tackle the static limit diversion of the susceptibility is a comprehensive analysis that includes imperfect holes' strings and final life times.

For the definition of $\bar{k}$ in [2], which is

$$\bar{k} = k - sign(k)2k_F, \qquad (17)$$

the maximum of the susceptibility for any $\omega$ occurs at $|q| = 2k_F$, with no dispersion. The definition in Eq.(17) is consistent with elastic NSE, **but is inconsistent with the low energy inward dispersions as revealed from inelastic NSE** [1, 3-6].

Consequently, the definition of $\bar{k}$ should be gradually changed as one is moving away for the Fermi level. The first suggestion that comes to mind is

$$\varepsilon_{\bar{k}} = -\varepsilon_k = Ja^2(k_F^2 - k^2) = Ja^2(\bar{k}^2 - k_F^2) \ , \qquad (18a)$$

together with the condition $sign(k) = -sign(\bar{k})$. Eq.(18a) yields immediately the relation

$$k^2 + \bar{k}^2 = 2k_F^2 . \qquad (18b)$$

Eq. (18b) suggests that the lengths of the vectors $k$ and $\bar{k}$ relate to each other as the two sides of a right angle triangle whose hypotenuse equals $\sqrt{2}k_F$. At the Fermi level $|k| = |\bar{k}| = k_F$, and when one of them tends to zero, the other one tends to $\sqrt{2}k_F$. Eq.(18a) has actually been used, for the approximation of $\varepsilon_{\bar{k}}$ at low energies throughout my former works [2, 12, 14]. However, here we suggest that when the wavevector $\bar{k}$ is to be specifically expressed, Eq. (17) should be replaced by Eq. (18b), and not only as a low energy approximation. In what follows we disregard the width of the susceptibility, and concentrate only on its peak at $\delta q = 0$. Then,

$$|q| = |\bar{k} - k| = \sqrt{k_F^2 + \varepsilon_+ / Ja^2} + \sqrt{k_F^2 - \varepsilon_+ / Ja^2} \ , \qquad (19)$$

which is approximated by

$$|q| \approx 2k_F(1 - \frac{\Lambda_-}{8J^2a^4k_F^4}\omega) \ . \qquad (20)$$

One good feature of the result is that the dispersion is inward. However, we show below that it is too small. At $k = 0$, $|\bar{k}| = |q| = \sqrt{2}k_F$, $\omega = \omega_r$, and we get

$$\frac{\omega_r}{J} = 8(1 - 2^{-1/2})(ak_F)^4 \frac{J}{|\Lambda_r|}. \tag{21a}$$

In Eq.(21a) $|\Lambda_r| = 2w_r v_r E_r \approx 2w_r v_r |\varepsilon_r| = 2w_r v_r J a^2 k_F^2$, which yields

$$\frac{\omega_r}{J} = 2.343 \frac{(ak_F)^2}{2w_r v_r}. \tag{21b}$$

We take a typical doping parameter for the under-doping regime $\delta = 1/8$, and get $ak_F = \pi/8$, and

$$\frac{\omega_r}{J} = \frac{0.36}{2w_r v_r}. \tag{21c}$$

The experimental results of Fig. 1 suggest that $\frac{\omega_r}{J} \approx 0.33$. For this to happen one needs to have $2w_k v_k \approx 1$. This is impossible since $2w_k v_k = 1$ at the Fermi level, where $k = k_F$. At $k = 0$, $2w_r v_r \ll 1$, and $\frac{\omega_r}{J} \gg 0.36$, contrary to experiment.

The dispersion of Eq. (20) is too small. Moreover, the minimum $|q|$, at the resonance frequency, is equal to $\sqrt{2}k_F$, rather than to zero. A second examination suggests that both of these faults stem from the momentum values that we have chosen for the particle-antiparticle pair. Inspired by the elastic experiments where these momentum values are $k$, and $\bar{k}$, we have continued the same perception to inelastic experiments. This perception is found to be misleading. A zero value for the minimum $|q|$ may be obtained if the nominal momenta of the pair are, $k$ and $-k$, so that $|q| = 2|k|$, goes to zero, when $|k|$ goes to zero. The advantage of this choice, in addition to the agreement with experiment in respect to reaching zero $|q|$, is the time reversal symmetry that is obtained between particle and anti-particle when their momenta are reversed in signs and equal in size. We also shall see that this choice yields approximately the correct dispersion. However, there is an intuitively conceptual difficulty in adopting this choice of momenta for the pair. The susceptibility $\Pi_+(q,\omega)$ stems from particle-antiparticle excitations. How could both the particle and the antiparticle carry the same nominal momenta (in absolute values)? The reader should keep in mind that the discussed momenta of the pair: $(k,-k)$ are only nominal, and the pair of excitations has also the components: $(\bar{k},-\bar{k})$. To clarify this we go back to the definition of the field $\psi_{k-}$ in Eq. (35a) of [2]

$$\psi_{k-}(t)=\begin{pmatrix}-w_k\\v_k\end{pmatrix}\gamma_{kA}\exp(-iE_{k-}t)+\begin{pmatrix}v_k\\-w_k\end{pmatrix}\eta_{kB}^{\dagger}\exp(iE_{k-}t) \tag{23a}$$

$$=\begin{pmatrix}w_k^2 & -w_kv_k\\-w_kv_k & v_k^2\end{pmatrix}\begin{pmatrix}C_k\\C_{\bar{k}}\end{pmatrix}_A\exp(-iE_{k-}t)+\begin{pmatrix}v_k^2 & -w_kv_k\\-w_kv_k & w_k^2\end{pmatrix}\begin{pmatrix}C_k\\C_{\bar{k}}\end{pmatrix}_B\exp(iE_{k-}t). \tag{23b}$$

The propagator of Eq. (6), which is originated from the field $\psi_{k-}$ includes two parts: a particle part, and an antiparticle part. The susceptibility $\Pi_+(q,\omega)$ is an integrated particle antiparticle product, where each of them has a nominal momentum. Notice that both the particle and the antiparticle parts of the propagators have the three components: $I$, $\tau_1$, and $\tau_3$, so that the I component of $\Pi_+(q,\omega)$, which results from $\tilde{G}_{ap}(k)\tilde{G}_p(-k)$ of Eq.(15), includes also the weighted products: $I \times I$, $\tau_1 \times \tau_1$, and $\tau_3 \times \tau_3$. The only diagonal component of $\Pi_+(q,\omega)$ is $I$, since its $\tau_3$ component vanishes. $\Pi_+(q,\omega)$ has also the off-diagonal components $\tau_1$ and $\tau_2$. Here we assume that the relevant component for our treatment is only the $I$ component. The reason for discarding the off-diagonal components is discussed in the end of the next section. Applying Eq.(15) for the particle anti-particle pair $(k,-k)$ yields,

$$\mathrm{Im}\,\Pi_+(2|k|,\omega)=\frac{|\Lambda_k|}{4Jak_F}\frac{[2w_k^2v_k^2I-w_kv_k\tau_1+iw_kv_k(w_k^2-v_k^2)\tau_2]}{\sqrt{\omega|\Lambda_k|}}\ . \tag{24}$$

Eq. (24) is valid only for small energies, namely when $\omega << \Lambda_k$, where its $I$ component converges into Eq.(16b). For a wider energy range, we go back to Eq. (14) and write

$$\Pi_+(q_\omega,\omega) = \frac{1}{8\pi\sqrt{J}} \int_{-Ja^2k_F^2+\delta}^{Ja^2k_F^2} \frac{d\varepsilon_k}{\sqrt{Ja^2k_F^2+\varepsilon_k}} \frac{\tilde{G}_{ap}(k)\tilde{G}_p(-k)}{\sqrt{\varepsilon_k^2+\Lambda_k^2}-|\Lambda_k|-\omega/2-i\delta} \,. \qquad (25)$$

In Eq. (25) $q_\omega = 2|k_\omega|$, where $k_\omega$ is determined by the pole of the integrand whose residue yields the integral. Changing the integration variable from $d\varepsilon_k$ to $dE_k$ yields

$$\Pi_+(q_\omega,\omega) = \frac{1}{8\pi\sqrt{J}} \int_{|\Lambda|+\lambda}^{E_{max}} \frac{dE_k E_k}{|\varepsilon_k|} [\frac{1}{\sqrt{Ja^2k_F^2-|\varepsilon_k|}} + \frac{1}{\sqrt{Ja^2k_F^2+|\varepsilon_k|}}] \frac{\tilde{G}_{ap}(k)\tilde{G}_p(-k)}{E_k-|\Lambda_k|-\omega/2-i\delta}$$

$$(26)$$

In Eq. (26), $E_{max} = \sqrt{J^2a^4k_F^4+\Lambda_r^2}$, and $\lambda$ is a small positive constant whose role is to exclude the $|\varepsilon_k|=0$ in the denominator of the integrand. Its integration, for $|\varepsilon_k| < 2Ja^2k_F^2$, yields approximately

$$\text{Im}\,\Pi_+(q_\omega,\omega) = \frac{1}{8Jak_F} \frac{\Lambda_\omega^2}{(\frac{\omega}{2}+|\Lambda_\omega|)\sqrt{\frac{\omega^2}{4}+\omega|\Lambda_\omega|}} \,. \qquad (27)$$

This result is conditional upon having a negative value for the polynomial in the denominator of the integrand of Eq.(26) at the lower limit of the integral, which happens when $2\lambda < \omega$. This way we avoid the divergence in the real static limit. Since $\lambda$ may be chosen very small, its introduction does not pose any shortcoming. When $\omega << 2|\Lambda_\omega|$, Eq.(27) converges into Eq.(16b). When $\omega >> 2|\Lambda_\omega|$, it approximates into

$$\text{Im}\,\Pi_+(q_\omega,\omega) = \frac{1}{8Jak_F} \frac{\Lambda_\omega^2}{(\frac{\omega}{2}+|\Lambda_\omega|)^2} \ . \tag{28}$$

The dispersion equation fits the poles of the integrand of Eq. (12), namely

$$\omega = 2(\sqrt{\varepsilon_k^2 + \Lambda_\omega^2} - |\Lambda_\omega|)\,. \tag{29}$$

For general $k$ we write: $|q_\omega| = 2\sqrt{k_F^2 + \varepsilon_k\,/\,Ja^2}$ , which together with Eq.(29) yields (for $k < k_F$)

$$|q_\omega| = 2k_F\sqrt{1 - \sqrt{\omega|\Lambda_\omega| + (\omega^2\,/\,4)}\,/\,Ja^2k_F^2}\ . \tag{30}$$

At $k = 0$, we assume that $|\Lambda_r| < Ja^2k_F^2$, and we approximately find that $\omega_r \approx 2Ja^2k_F^2 - 2|\Lambda_r| \approx 2Ja^2k_F^2$. For $\delta = 1/8$, $ak_F = \pi/8$, and we get $\omega_r\,/\,J \approx \pi^2\,/\,32 \approx 0.31$. **This result is very close to the value suggested by the results in Fig. 1.** For $\omega < 2Ja^2k_F^2$, which is correct almost throughout the whole energy range, we approximate

$$|q_\omega| = 2k_F[1 - \sqrt{(\omega|\Lambda_\omega| + \frac{\omega^2}{4})\,/\,2Ja^2k_F^2}\,]\ \ . \tag{31a}$$

The essential difference between Eq. (31a), and Eq. (20) stems from the cancellation of the linear part of $\varepsilon_+$ in Eq. (19), which does not occur in Eq. (31a). For still lower energy, $\omega << |\Lambda_\omega|$, we write

$$|q_\omega| \approx 2k_F[1 - \sqrt{\omega\Lambda_\omega}\,/\,2Ja^2k_F^2]\ . \tag{31b}$$

Eq. (31b) suggests that by the Fermi momentum, $\omega \propto (|q| - 2k_F)^2$, rather than linear in $(|q| - 2k_F)$. However, it approaches $\omega_r$ linearly in $|q_\omega|$.

## 3. Comparison with Experiment and conclusions

The dispersion curve in Fig. 1, which is based on several experimental studies, is undoubtedly unusual, even when we consider only its low energy regime. In general, dispersion curves are classified into two groups: 1) Acoustical- which starts from zero energy and momentum, and rises outwardly (with positive slop). 2) Optical- which starts from finite energy and may go up or down with increasing momentum, but usually do not end at zero energy [15]. The curve in Fig. 1 fits none of these two groups. Its zero energy at $|q| = 2k_F$ seems to be essential. The present paper provides a theoretical solution to the experimental data. The given model provides two functions, one is the dispersion curve, and the other is the imaginary part of the susceptibility. The dispersion curve is given for two possible definitions of $\bar{k}(k)$, Eq. (17), and Eq. (18b). It has been proven that none of these definitions results in the experimental dispersion when the pair momenta are $(k, \bar{k})$. On the other hand none could be discarded when the pair momenta are $(k, -k)$.

So far we have dealt only with the low energy part of the hour-glass dispersion. The high energy part ($\omega > \omega_r$) of the hour-glass dispersion is less conclusive. This is so because the high energy part may involve more mechanisms, which complicate the analysis. Such one possible mechanism may involve spin waves as in the mother AF materials, which are not originated by movements of strings of holes, but rather by spins flip due to the anti-ferromagnetic coupling of neighboring spins. The self sustained incommensurate spin waves in the condensate that are produced by the movement of the strings of holes, suggest also the production of spin waves excitations due to the $H_J$ part of the $H_{tJ}$ Hamiltonian. For a single $CuO_2$ layer, these $H_J$- originated spin waves are of acoustical nature, as has been observed for $La_2CuO_4$ and other mother AF materials [1, 16]. This implies that, for a given low energy, $\omega << \omega_r$, the momenta of the $H_J$ originated spin waves are smaller than their holes driven counterparts, and consequently, their associated (phase) velocity is larger. This in turn, suggests that their propagation is stalled at the strings of holes, where they may be multiply reflected and quenched. This is also the situation when the string of holes is taken as a reflecting boundary where the neighboring spin interaction is missing. The long wavelength SDW that is reflected at this boundary is inversed in phase with respect to the incoming SDW, and cancels it. This explains the lack of observation of low energy spin waves with momenta around $|q| = \omega |v|^{-1} \approx Q_{AF}$. We assume that in the low energy range the holes driven spin waves dominate the $H_J$ driven spin waves. We also assume that when the energy approaches the resonance energy from below, the phase velocities of the two waves become comparable and they might mix. From this energy and up the discussion is only qualitative. This mixture is out of the scope of the present theory, and therefore the susceptibility intensities close to the resonance energy, which have been evaluated in former paragraphs, are not valid. The high energy outwardly dispersed waves may be a mixture of $H_J$ driven spin waves with holes driven waves whose energy is out of the range of the condensate..

The above discussion puts one in perspective with respect to the comparison between theory and experiment, which implies the following: 1) We consider only the low energy inwardly dispersing part of Fig.1. 2) The part of the YBCO sample is also removed due to limited range and data points. Thus, we are left with data of the samples: $La_{1.90}Sr_{0.10}CuO_4$ , $La_{1.875}Ba_{0.125}CuO_4$, and $La_{1.84}Sr_{0.16}CuO_4$[1,4,6]. For these samples, we plot Eq.(30) for the dispersion relations. We assume an arbitrary energy dependent pseudogap according to: $|\Lambda_\omega| = |\Lambda| / (1 + \dfrac{\omega^2}{\Lambda^2})$. The only fitted parameters are $|\Lambda| = J / 20$ (which is roughly 7.3mV for LSCO), and $\delta$. For $La_{1.875}Ba_{0.125}CuO_4$, we took $\delta = 0.125$; for $La_{1.90}Sr_{0.10}CuO_4$, $\delta = 0.12$; and for $La_{1.84}Sr_{0.16}CuO_4$, $\delta = 0.145$. The latter value is justifiable by the saturation of $\delta(x)$, close to the optimal doping [1,25]. For the large value of $La_{1.90}Sr_{0.10}CuO_4$, we have no justification except for a better fit to experiment. The dispersion relations for the discussed samples are shown in Fig. 2. We remark that the effect of the fitted parameter $|\Lambda|$ is not dramatic, as long as it is kept smaller than $\omega_r$. This means that the theoretical dispersion curves are almost essential, in the sense that they are not artificially manipulated by unphysical fitted parameters.

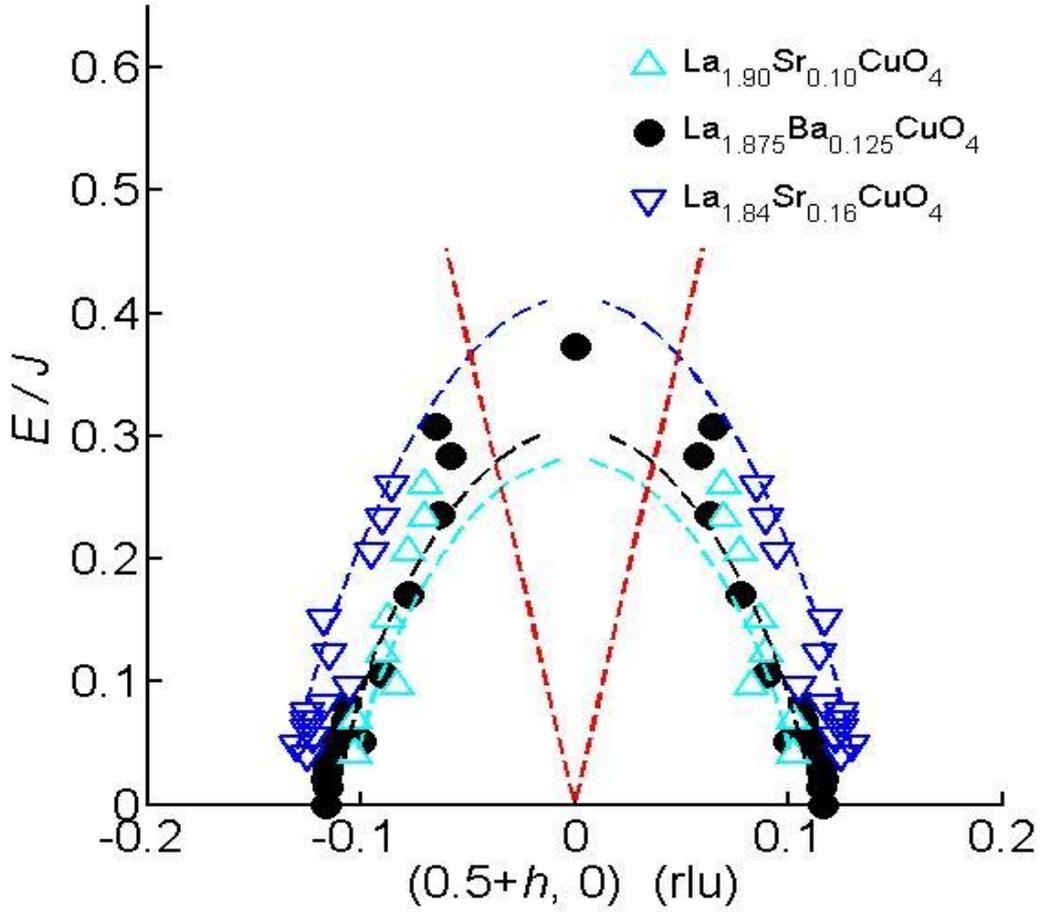

**Figure 2.** The LSCO and LBCO low energy data points of Fig. 1 are plotted together with the dispersion relations of Eq.(30), which are shown by dashed lines of the corresponding colors. The red straight lines depict the dispersion of spin waves (in the x and y directions) in the undoped material $La_2CuO_4$.

From Fig. 2 we immediately see that the curves of the different samples do not converge to the same normalized energy as $q$ goes to zero. This is in accord with the former section where it is shown that $\omega_r \approx 2Jk_F^2a^2 \approx 2J(\pi\delta)^2$. The fit to $La_{1.84}Sr_{0.16}CuO_4$ is quite good, but the fit to the other two samples is good only at energies lower than $\approx 0.24J$. I speculate that the departure from experiment at these higher energies is due to the interference, of the $H_{tJ}$ originated SDW, with the $H_J$ originated SDW. In order to illustrate this, the dispersion of the latter was duplicated from [1, 16], and is shown by the red dashed lines in Fig. 2. We see that, for the $La_{1.84}Sr_{0.16}CuO_4$ sample, the theoretical curves intersect with the direct spin flip curve roughly at $E/J \approx 0.37$, much higher than the highest energy of the experimental data (which is $E/J \approx 0.26$). On the other hand, the data points of the underdoped samples, which deviate from the theoretical curves, are either above the intersection points, or close to it from below. This suggests that the $H_{tJ}$ originated SDW interfere with the direct spin flip SDW, for the shown data of the underdoped samples, and therefore shift their dispersion. It does so to a much lesser extent for $La_{1.84}Sr_{0.16}CuO_4$. This same interference mechanism might also explain the disagreement between the observed intensities and the lower intensities predicted by the present theory at high energies.

The above comparison with experiment has been done only for the dispersion. For comparison of the susceptibility intensities with experiment we consider the works of Cristensen et al. [4], of Tranquada et al. [5], and of Hinkov et al. [15]. Even a preliminary comparison of the susceptibility to experiment indicates difficulties. When attempting the comparison, one has to note that the present treatment, being two dimensional in the propagator matrix, **excludes** various effects of superconductivity [1, 3-6, 12]. These effects include spin gap and its associated magnified intensity at the spin edge and above it [3-5]. Thus, one is not sure whether the experimental susceptibility reduction with increasing energy is originated from Eq.(27), or is the result of getting farther from the gap edge. This difficulty together with the experimental error, make a quantitative comparison ineffective. Moreover, the perfect strings of holes on which our model is based, and the lack of assuming finite life times for the strings as well as for the excitations, undoubtedly distorts the energy dependence of the susceptibility. This has already been demonstrated by the simplified model which led to Eq.(16c). Additional important source for the uncertainty of Eq.(27) and Eq.(30) is the possible interference of spin waves that are originated by direct spin flip, especially at energies close to $\omega_r$. This has already been discussed above in the present section. Thus, Eq.(27) has to be considered only as the simplified approximation to the more realistic susceptibility. All one can claim at this stage is that the susceptibility decrease with increasing energy beyond the spin gap is a feature that is common to the above mentioned experiments and Eq.(27).

The above comment about the difficulties that are encountered when attempting a comparison of the intensity with experiment is only part of the problem. More difficulties of the same kind are discussed bellow. With the choice of $(k, -k)$ as the pair momenta, the resulted intensity is supposedly weak at $|q| = 0$, compared with the intensity at $|q| = 2k_F$. Their ratio is roughly equal to $\sqrt{\dfrac{2\Gamma}{\Lambda}}(1 + \dfrac{\omega_r}{2|\Lambda_r|})^{-2}$. If one assumes that $\omega_r \approx 2Ja^2k_F^2 >> 2|\Lambda_r|$, then the ratio is roughly equal to $\sqrt{\dfrac{2\Gamma}{\Lambda}}(\dfrac{2|\Lambda_r|}{\omega_r})^2$.

Thus, $|\Lambda|$ has a profound effect on the intensities. This is contrary to the dispersion for which $|\Lambda|$ has no dramatic effect as long as it is kept within the reasonable limits. For the underdoped $YBCO$, the resonance energy is $\omega_r = 34 meV$ for the $YBa_2Cu_3O_{6.6}$ sample of Hayden et al. [7], and $\omega_r = 33 meV$ for the $YBa_2Cu_3O_{6.5}$ sample of Stock et al. [8]. The determination of the pseudogap is more uncertain, despite the many investigations that studied the pseudogap by means of many experimental tools. This is so due to uncertainties in interpreting the experimental data. The results of many experimental tools were interpreted to be related to a pseudogap, or to a "gap in the normal state". Such tools are ARPES, tunneling, NMR, dc conductivity and optical conductivity, as well as thermodynamic properties [17]. The first problem of interpretation stems from not discriminating between single particles (electrons or holes) and condensed strings of holes. This is apparent especially in ARPES and tunneling where a single electron is extracted from the cuprate or injected into it. Note that contrary to regular superconductors where removing electron demands energy of $2\Delta$ to break a Cooper pair, here a single string is made of a huge number of holes, and changing their number by one does not necessarily break it. The common practice in ARPES or tunneling is to measure differential currents, as a function of energy and temperature, and to observe a gap which persists when the temperature is raised above the superconductive $T_c$ [17-24]. Fitting this to our model where the pseudogap energy is related to a string of a large number of holes is questionable. However, I proposed in [12] a model in which the superconductive state is made of a superposition of terms where part of them is based on perfect strings, and part on imperfect strings, with either access or deficiency of holes. These strings exchange holes between each other in a correlated fashion to result in the known superconductive features, such as small weight in the gap region and large weight by the gap edge. When the temperature is increased above the superconductive $T_c$, imperfect strings still exist in part, due to thermal fluctuations. However, since now their exchange is not coherently correlated the maxima by the gap edge is lost. The single electrons that are removed from the system or injected to it are inherently related to these imperfect strings, which are subject to a pseudogap. When a single electron is removed from the system the initial energy is $NE_F^N$, where $E_F^N$ is the chemical potential of a system with N condensed holes, and the final energy is $(N+1)E_F^{N+1}$. The difference in energy is: $N(E_F^{N+1} - E_F^N) + E_F^{N+1}$. To relate this difference to the pseudogap the theory needs further development- to include thermal fluctuations and imperfect strings. Anyway, strong efforts to calculate intensities at energies close to $\omega_r$ are futile because of the interference of the modes that are originated from direct spin flip, as discussed above.

The advantage of choosing the particle anti-particle momenta to be **k** and **–k** enables full control of $|q|$ by the continuous change of $|k|$. When k is scanned between zero and $k_F$, $|q|$ is changed from zero to $2k_F$, and $\omega$ - from $\omega_r$ to zero. When k is further increased beyond $k_F$, one can repeat the same treatment using $\bar{k}$ (which absolute value is now smaller than $k_F$) instead of $k$. Since $E_k = E_{\bar{k}}$, and $|\Lambda_k| = |\Lambda_{\bar{k}}|$, the equations involving energy remain the same. However, $|q_\omega| = 2|k|$ now is out of the momentum range $(-2k_F, 2k_F)$, and one should apply the rule $|q_\omega| = 2|\bar{k}|$. This is justifiable by the rule that the system conserves momentum only modulus $(k - \bar{k})$ [26]. Notice that $(k - \bar{k})$ is a constant independent of $k$, when $\bar{k}(k)$ is defined by Eq.(17), but is dependent upon $k$, when it is defined by Eq. (18b). In both definitions $|q_\omega| = 2|\bar{k}|$ is scanned between $2k_F$ and zero when $k$ is scanned between $k_F$ and its upper limit, producing the same dispersion curve as the one obtained by scanning through $k < k_F$. Although none of the said definitions of $\bar{k}(k)$ should be completely ruled out without further investigation, the definition of Eq.(17) seems to be more in accord with the observed sharp incommensurate peaks at zero energy [1, 25].

A comparison between the present investigation and the results of [2] is needed. The perception there was that the particle anti-particle pair $(k, \bar{k})$ is the one which originates the static SDW, whereas here the perception is that the pair is $(k, -k)$. We notice that at the Fermi level $\bar{k} = -k$. However, the other terms in the sum of Eq.(25) of [2] do not equate for the two different perceptions. Moreover, the time dependence is not given in Eq.(25) of [2]. Anyway, the static limit should be reanalyzed, after incorporating effects of life times and imperfect strings of holes.

Let us now discuss the off-diagonal components of Eq.(24) and Eq.(25), which result from $\tilde{G}_{ap}(k)\tilde{G}_p(-k) = [2w_k^2 v_k^2 I - w_k v_k \tau_1 + i(w_k^2 - v_k^2)\tau_2]$. The $\tau_2$ component of this product is not commutative, and changes sign under the exchange of the order of the $\tilde{G}'s$. While the elements of the $\tau_1$ component are equal to each other, the elements of the $\tau_2$ component break the asymmetry between the two off-diagonal elements of the matrix. In any case these two elements are negative, which disqualifies them as elements of scattering cross section. Another aspect of these off-diagonal elements is their momentum. While the momentum of the diagonal component is $|q| = |k - (-K)| = |2k|$, the momentum of the off-diagonal components is shifted by $2k_F$ [26], which should make their low energy dispersion (if existed) to be acoustic, starting at $q = 0, \omega = 0$. However, they are theoretically forbidden, and have never been observed experimentally.

To summarize we argue that, despite some difficulties, the perception that the origin of the pseudogap is attributed to a condensate of linear strings of holes is further established in the present analysis. The model that was introduced first in [2], were it was shown that it is the "natural" model to account for the observed static spin waves, has been further extended here to include the inelastic waves, and their unusual dispersion relations. On the way we have pointed out the necessary improvements that are still needed to promote the model to its maturity.